\documentclass[pra,twocolumn,superscriptaddress,amsmath,amssymb,showpacs]{revtex4}

\usepackage{graphicx}
\usepackage{dcolumn}
\usepackage{bm}
\usepackage[bookmarks=true]{hyperref}
\usepackage[usenames,dvips]{color} 

\newcommand{\ket}[1]{|#1\rangle}
\newcommand{\bra}[1]{\langle #1|}
\newcommand{\tr}{\ensuremath{{\rm Tr}}}
\newcommand{\ham}{\mathcal{H}}
\begin{document}

\title{NMR multiple quantum coherences in quasi-one-dimensional
spin systems: \\ Comparison with ideal spin-chain dynamics}

\author{Wenxian Zhang} \affiliation{Department of Physics and
Astronomy, Dartmouth College, Hanover, New Hampshire 03755, USA}
\affiliation{Department of Optical Science and Engineering, Fudan
University, Shanghai 200433, China}

\author{Paola Cappellaro} \affiliation{ITAMP Harvard-Smithsonian
Center for Astrophysics, Cambridge, Massachusetts 02138, USA}

\author{Natania Antler} \affiliation{\mbox{Department of Physics, 
Massachusetts Institute of Technology,
Cambridge, Massachusetts 02139, USA}}

\author{Brian Pepper} \affiliation{\mbox{Department of Physics, 
Massachusetts Institute of Technology, Cambridge,
Massachusetts 02139, USA}}

\author{\mbox{David G. Cory}} \affiliation{\mbox{Department of Nuclear
Science and Engineering, Massachusetts Institute of Technology,
Cambridge, Massachusetts 02139, USA}}

\author{Viatcheslav V. Dobrovitski} \affiliation{Ames Laboratory, US DOE, Iowa
State University, Ames, Iowa 50011, USA}

\author{Chandrasekhar Ramanathan} \affiliation{\mbox{Department of
Nuclear Science and Engineering, Massachusetts Institute of
Technology, Cambridge, Massachusetts 02139, USA}}

\author{Lorenza Viola} 
\thanks{Email address: Lorenza.Viola@Dartmouth.edu}
\affiliation{Department of Physics and
Astronomy, Dartmouth College, Hanover, New Hampshire 03755, USA} 

\date{\today}

\begin{abstract}
The $^{19}$F spins in a crystal of fluorapatite have often been used
to experimentally approximate a one-dimensional spin system.  Under
suitable multi-pulse control, the nuclear spin dynamics may be modeled
to first approximation by a double-quantum one-dimensional
Hamiltonian, which is analytically solvable for nearest-neighbor
couplings. Here, we use solid-state nuclear magnetic resonance
techniques to investigate the multiple quantum coherence dynamics of
fluorapatite, with an emphasis on understanding the region of validity
for such a simplified picture.  Using experimental, numerical, and
analytical methods, we explore the effects of long-range intra-chain
couplings, cross-chain couplings, as well as couplings to a spin environment,
all of which tend to damp the oscillations of the multiple quantum
coherence signal at sufficiently long times.  Our analysis
characterizes the extent to which fluorapatite can faithfully simulate
a one-dimensional quantum wire.
\end{abstract}


\maketitle


\section{Introduction}

Low-dimensional quantum spin systems are the subject of intense
theoretical and experimental investigation. From a condensed matter
perspective, not only do these systems provide a natural setting for
deepening the exploration of many-body quantum coherence properties as
demanded by emerging developments in spintronics and nanodevices
\cite{Awschalom07,Childress06,Dutt07}, but the ground states of
one-dimensional (1D) conductors provide insight into the solution of
the one-band Hubbard Hamiltonian \cite{Essler05}.  From a quantum
information perspective \cite{Nielsen00}, quantum spin chains have
been proposed as quantum wires for short-distance quantum
communication, their internal dynamics providing the mechanism to
coherently transfer quantum information from one region of a quantum
computer to another \cite{Bose03} (see also \cite{Bose08} for a recent
overview).  Perfect state transfer, in particular, has been shown to
be theoretically possible by carefully engineering the couplings of
the underlying spin Hamiltonian.  A number of efforts are underway to
devise protocols able to achieve reliable quantum information transfer
under more realistic conditions -- bypassing, for instance, the need
for initialization in a known pure state \cite{DiFranco08}, explicitly
incorporating the effect of long-range couplings
\cite{Kay06,Avellino06,Gualdi08}, or exploiting access to external end
gates \cite{Burg06,Zhang09}.  Still, few (if any) physical systems can
meet the required constraints, and it is likely that quantum
simulators will be needed to experimentally implement these schemes.
Of course, quantum simulators will in turn allow us to probe a much
broader range of questions encompassing both quantum information and
condensed matter physics \cite{Lloyd96}.  Optical lattices have shown
much promise in simulating quantum spin systems \cite{Bloch08}.  Among
solid-state devices, coupled spins in apatites have recently enabled
experimental studies of 1D transport and decoherence dynamics
\cite{Cho06,Cappellaro07a,Cappellaro07b,Oliva08}.  

Fluorapatite (FAp) has long been used as a quasi-1D system of nuclear
spins.  Lowe and co-workers characterized the nuclear magnetic
resonance (NMR) line shape of FAp \cite{Engelsberg73,Sur75a}, and
described the dipolar dynamics of the free induction decay in terms of
the 1D XY model \cite{Sur75b}.  Cho and Yesinowski 
investigated the many-body dynamics of FAp under an effective
double-quantum (DQ) Hamiltonian, and showed that the growth of
high-order quantum coherences was distinctly different from that
obtained in dense 3D crystals \cite{Cho93,Cho96}. From a theoretical
standpoint, FAp provides a rich testbed to explore the controlled time
evolution of a many-body quantum spin system. The DQ Hamiltonian is
analytically solvable in the tight-binding limit, where only nearest
neighbor (NN) couplings are present
\cite{Feldman97,Doronin00,Cappellaro07a}.  Previous work showed that
the implementation of a DQ Hamiltonian in the FAp system using
coherent averaging techniques is a promising tool for the study of
transport in quantum spin chains.  We demonstrated, in particular,
that the DQ Hamiltonian is related to the XY-Heisenberg Hamiltonian by
a similarity transformation, and that it is possible to transfer
polarization from one end of the chain to the other under the DQ
Hamiltonian \cite{Cappellaro07b}.  In fact, the signature of this
transport shows up in the collective multiple quantum coherence (MQC)
intensity of the spin chain.  Experimentally, it has also been shown
that it is possible to prepare the spin system in an initial state in
which the polarization is localized at the ends of the spin chain
\cite{Cappellaro07a}, paving the way towards achieving universal
quantum control \cite{Fitzsimons06}.

Since the mapping between the experimental system and the idealized
model \cite{Cappellaro07a,Cappellaro07b} is not perfect, an essential
step forward is to address where and how this model breaks down, which
constitutes the main aim of this paper.  In particular, we
systematically examine the viability of using NMR investigations of
FAp as a test-bed for 1D transport, by relying on a combination of
experimental and numerical methods.  We first examine the effects on the
relevant observables of experimental
errors introduced during the implementation of the DQ Hamiltonian,
which arise due to higher-order terms in the average Hamiltonian
describing the effective spin evolution.  We also examine errors
introduced in some state initialization sequences due to the
restriction of the control fields to collective rotations.  Since the
FAp crystal is in reality a three-dimensional (3D) lattice, we next
investigate in detail how the spin dynamics is affected by the
presence of longer-range couplings, both within a single chain and
between adjacent spin chains.

The content of the paper is organized as follows. We describe the
quasi-1D spin system of FAp in Sec.~\ref{sec:sys}, including the
evolution in the absence of control as well as the dynamics under
suitable pulse sequences.  In the same section, we also discuss the
system initialization and the readout of the experimental MQC
signal. Sections~\ref{sec:mqc} and \ref{sec:beyond} present both
experimental and numerical results of MQC dynamics, and are the core
of the paper.  By comparing the numerical results with the analytical
predictions available in the limiting case of a DQ Hamiltonian with NN
couplings, we evaluate the effect of high-order average Hamiltonian
terms, next-nearest-neighbor (NNN) couplings, and cross-chain
couplings between multiple chains.  Our findings are summarized in
Sec.~\ref{sec:con}.  Appendix \ref{sec:num} presents technical
background on the relevant numerical methodology, whereas we also
include in Appendix \ref{sec:mirror} a description of finite size
effects as found in simulations, and in Appendix \ref{sec:ChBath} a
discussion of an alternative chaotic model for the spin bath.


\section{Physical system and experimental settings}
\label{sec:sys}

\subsection{Spin Hamiltonian of fluorapatite}

We consider a single crystal of FAp [Ca$_5$(PO$_4$)$_3$F] at room
temperature, placed in a strong external magnetic field along the
$z$-direction that provides the quantization axis for the nuclear
spins. It is possible to truncate the magnetic dipolar interaction
among the spins in this strong field, keeping only the secular terms.
The resulting secular dipolar interaction \cite{Slichter92} among $N$
$^{19}$F nuclear spin-1/2 is anisotropic due to the presence of the
quantization field, leading to a Hamiltonian of the form:
\begin{eqnarray}
{\cal H}_{dip} &=& \sum_{j<\ell}^N b_{j\ell}
\left[\sigma^z_j\sigma^z_{\ell} - \frac{1}{2}
(\sigma^x_j\sigma^x_{\ell}+\sigma^y_j\sigma^y_{\ell})\right].
\label{eq:dip}
\end{eqnarray}
Here, $\sigma_j^\alpha$ ($\alpha = x,y,z$) denotes the Pauli
matrices of the $j$th spin and $b_{j\ell} = (\mu_0/16\pi)
(\gamma^2 \hbar /r_{j\ell}^3) (1-3\cos^2\theta_{j\ell})$, with
$\mu_0$ the standard magnetic constant, $\gamma$  the gyromagnetic
ratio of fluorine, $r_{j\ell}$ the distance between nucleus $j$
and $\ell$, and $\theta_{j\ell}$ the angle between $\vec
r_{j\ell}$ and the $z$-axis. The geometry of the spin system is
reflected in the distribution of the $b_{j\ell}$ couplings.

\begin{figure}[htb]
\centering \includegraphics[width=2.8in]{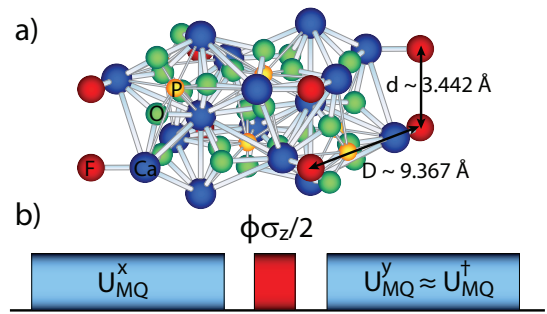}
\caption{a) Unit cell of the fluorapatite crystal
[Ca$_5$(PO$_4$)$_3$F], highlighting the geometry of the fluorine
chains (in red). b) NMR pulse sequence for the creation and detection
of MQC.}
\label{fig1:FApPulseSeq}
\end{figure}

The FAp crystal has a hexagonal geometry with space group
P$6_3$/m~\cite{Lugt64} (see Fig. \ref{fig1:FApPulseSeq}.a). The
dimensions of the unit cell are $D=9.367$~\AA\;and $c=6.884$~\AA. The
$^{19}$F nuclei form linear chains along the $c$-axis, each one
surrounded by six other chains. The distance between two intra-chain
$^{19}$F nuclei is $d=c/2=3.442$~\AA \; and the distance between two
cross-chain $^{19}$F nuclei is $D$. The largest ratio between the
strongest intra- and cross- chain couplings ($\approx 40$) is obtained
when the crystalline $c$-axis is oriented parallel to the external
field. Thus, to a first approximation, in this crystal orientation the
3D $^{19}$F system may be treated as a collection of many identical 1D
chains.  For a single chain oriented along $z$, we have $b_{j\ell} =
-(\mu_0/\pi) (\gamma^2 \hbar/c^3|j-\ell|^3)$.

In reality, naturally occurring defects in the sample (such as
vacancies or substitutions~\cite{FapDefects}) cause the chains to be
broken into many shorter chains.  Here we model the system as
 an ensemble of (approximately) {\em independent and
equivalent chains} with finite length. Such a simplified description 
is necessary to obtain a computationally tractable model.

\subsection{Control capabilities and effective dynamics}

\subsubsection{Unitary control}
Unitary control is obtained by applying (near)resonant radio-frequency 
(rf) pulses to the spin system. FAp contains $^{19}$F and $^{31}$P 
spins-$1/2$, both of which are 100\% abundant. Moreover, in an ideal 
crystal, all the $^{19}$F spins are chemically equivalent, as are all 
the $^{31}$P spins. As a consequence, all rf control pulses are applied 
{\em collectively} to all the spins.

In NMR, the term MQC refers to coherences between two or more spins.
When the system is quantized along the $z$-axis, a quantum coherence
of order $n$ is associated to the transition between two states
$\ket{m_1}$ and $\ket{m_2}$, such that the difference of the magnetic
moment along $z$ of these states $(m_1-m_2)\propto n$. That is,
multiple quantum coherences of order $n$ describe states like
$\ket{m_2}\bra{m_1}$, or elements in the density matrix that
correspond to a transition between these two states \cite{MQCbook}.
Quantum coherences can also be classified based on their response to a
rotation around the $z$ (quantization) axis. A state of coherence
order $n$ acquires a phase proportional to $n$ under a $z$-rotation.
Multiple quantum NMR techniques
\cite{Hatanaka75,Pines76,Aue76,Vega76,Warren80} have enabled
researchers to probe multi-spin processes, and gain insight into the
many-body spin dynamics of dipolar-coupled solids
\cite{Baum85,Munowitz87b,Levy92,Lacelle93,Ramanathan03,Cho05,Cho06}.

To study the MQC dynamics of the spin system, we typically let it
evolve under the DQ Hamiltonian
\begin{eqnarray}
{\cal H}_{DQ} &= &\sum_{j <\ell} \frac{b_{j\ell}}{2}
(\sigma^x_j\sigma^x_{\ell}-\sigma^y_j\sigma^y_{\ell}) \nonumber \\
&=& \sum_{j <
\ell} b_{j\ell} (\sigma^+_j\sigma^+_{\ell}+\sigma^-_j\sigma^-_{\ell}),
\label{eq:dq}
\end{eqnarray}
with $\sigma_j^\pm = (\sigma_j^x \pm i\sigma_j^y)/2$.  Following
\cite{Baum85,Ramanathan03}, we utilize a $16$-pulse cycle applied
on-resonance with the $^{19}$F Larmor frequency to implement the DQ
Hamiltonian to lowest order in AHT
description~\cite{Slichter92,Haeberlen76,AHTnote}.  A key feature of
this sequence is that the fluorine-phosphorus dipolar interaction is
decoupled, which makes it possible to ignore the presence of the
$^{31}$P spins in the rest of this paper. The dynamics during the
pulse sequence can be written in terms of an effective Hamiltonian
$\overline{\ham}_{DQ}$,
\begin{eqnarray}
    U^{x(y)}_{MQ}(t)&=&\mathcal{T}\exp\left(-i\int_0^t [\ham_{dip}
    + {\cal H}^{x(y)}_{rf}(s)]ds\right)\nonumber  \\
    &=& e^{\pm i\overline{\ham}_{DQ}t},
\label{MQpropagator}
\end{eqnarray}
where $\mathcal{T}$ denotes time-ordering operator, $\hbar=1$, and
${\cal H}_{rf}^{x (y)}(t)$ is the time-dependent Hamiltonian
describing the rf-pulses along the $x$- (or $y$-) axis (whereby the
corresponding $\pm$ sign in front of the effective Hamiltonian).

\subsubsection{Initialization capabilities}

The spin dynamics under the DQ Hamiltonian depends critically on the
initial state in which the system is prepared.  Here, we focus our
attention on two choices of direct experimental relevance
\cite{Cappellaro07a}. One is the equilibrium {\em Zeeman thermal
state}, which is obtained at the thermal
equilibrium in a strong external magnetic field ($B_0=7$ T in our
experiments) at room temperature. The thermal state can be expressed
as
\begin{eqnarray}
\rho'_{th}(0) &\propto & \exp(-\varepsilon \sigma_z)\approx \openone -
\varepsilon \sigma_z,
\label{eq:ts}
\end{eqnarray}
where $\sigma_z = \sum_j \sigma_j^z$ and $\varepsilon = \gamma B_0 /
k_B T$, with $k_B$ the Boltzmann constant and $T$ the temperature
($\varepsilon \approx 10^{-5}$ at room temperature for FAp).  In line
with standard NMR practice, we consider only the evolution due to the
component proportional to $\varepsilon$, $\rho_{th}(0) = \sigma_z$,
since the identity matrix does not evolve or contribute to the MQC
signal under the assumption of unital dynamics.  The second initial
state that is experimentally available is a mixture of states where
only a spin at the extremities of the chain is polarized, which can be
{\em formally} represented as
\begin{eqnarray}
\rho_{end}(0) &=& \sigma^z_1+\sigma^z_N,
\label{eq:endp}
\end{eqnarray}
where spin 1 and spin $N$ are located at the two ends of the spin
chain~\cite{Cappellaro07a}.  We refer to this as the {\em
end-polarized state}.  A description of the method used to create this
state is given in Sec.~\ref{sec:endprep}.

\subsubsection{Readout capabilities}

In an inductively detected NMR experiment (in which a coil is used to
measure the average magnetization), the observed signal is $S(t)=\zeta
\langle \sigma^{-}(t) \rangle =\zeta
\mathrm{Tr}\left\{\sigma^{-}\rho(t)\right\}$, where $\sigma^{-} =
\sum_j \sigma_{j}^-$ and $\zeta$ is a proportionality constant. The
only terms in $\rho(t)$ that yield a non-zero trace, and therefore
contribute to $S(t)$, are angular momentum operators such as
$\sigma_{j}^+$, which are single-spin, single-quantum
coherences. Thus, in order to characterize multi-spin dynamics, it is
necessary to indirectly encode the signature of the dynamics into the
above signal. This is precisely what is done in standard NMR MQ
spectroscopy, using an evolution-reversal experiment
\cite{MQCbook}. The density operator at the end of an MQ experiment is
given by
\begin{equation}
\rho_f = U_{MQ}^\dag U_{\textrm{evol}} U_{MQ} \rho_i U_{MQ}^{\dag}
U_{\textrm{evol}}^\dag U_{MQ},
\end{equation}
where $U_{MQ} = \exp(-i {\cal H}_{DQ}t)$, and $U_{\textrm{evol}}$
determines the nature of the information encoded (see
Fig. \ref{fig1:FApPulseSeq}.b). 

In our experiment, we are interested in the evolution of MQC under the
DQ Hamiltonian, thus we measure the signal as we systematically
increase $t$.  In order to encode information about the distribution
of the MQC, we apply a collective rotation about the $z$ quantization
axis, $U_{\textrm{evol}} = \exp(-i\phi\sigma_z/2)$. Then, to extract
the coherence order distribution, the measurement is repeated while
incrementing $\phi$ from 0 to $2\pi$, in steps of $\delta\phi =
2\pi/2K$, where $K$ is the highest order of MQC encoded. The signal acquired 
in the $k$-th measurement is
then $S_z^k(t)=\tr[\rho^k(t) \sigma_z]$, where $\rho^{k}(t)$ is the
density matrix evolved under the propagator
$U_k(t)=e^{i\overline{\ham}_{DQ}t} e^{-i
k\delta\phi\sigma_z/2}e^{-i\overline{\ham}_{DQ}t}$, and we have
assumed that $\sigma_z$ is the experimental observable. In practice,
we use either a $\pi/2$ pulse or a solid echo \cite{SolidEcho} to read
out the signal at the end of the experiment.  Fourier-transforming the
output with respect to $\phi$ yields the coherence order intensity:
\begin{eqnarray}
J_n(t) &=& \sum_{k=1}^{K} S_z^k(t) e^{-ikn\delta\phi}. \label{eq:Jn}
\end{eqnarray}
Note that since the initial states we consider are population terms in
the $z$-basis, the final states at the end of the evolution-reversal
experiment are also population terms (hence our use of the observable
$\sigma_z$).


\section{Multiple quantum dynamics: Simple model and experiment}
\label{sec:mqc}

\subsection{Ideal spin-chain dynamics}

The fact that the evolution of the 1D spin chain under a DQ
Hamiltonian is exactly solvable in the tight binding limit
\cite{Feldman97,Doronin00,Cappellaro07a} provides a useful starting
point for theoretical analysis.  Hereafter, we shall refer to this
model as the analytical model.  Moreover, the DQ Hamiltonian is
related to the XY Hamiltonian by a similarity transformation that
inverts every alternate spin in the chain
\cite{Doronin00,Cappellaro07b}. Besides using the analytical results
to calibrate our numerical methods (see Appendix~\ref{sec:num}), we
will also investigate the effect of long-range interactions beyond the
NN limit, by comparing numerical and analytical results. For
convenience, we set the NN coupling strength in the DQ Hamiltonian of
Eq.~(\ref{eq:dq}) $b\equiv b_{12}=1$, so that time shall be measured
in units of $1/b$ henceforth (unless explicitly stated otherwise).

For both the thermal and the end-polarized initial state, only zero
and DQ coherences are predicted by the analytical model. Specifically,
for the thermal initial state, the normalized intensities are
\begin{eqnarray}
J_0^{th}(t) &=& \frac{1}{N} \sum_k \cos^2(4bt\cos\psi_k), \nonumber \\
J_2^{th}(t) &=& \frac{1}{2N} \sum_k \sin^2(4bt\cos\psi_k),
\label{eq:j02th}
\end{eqnarray}
where as before $N$ is the number of spins in the chain and
$\psi_k=k\pi/(N+1)$. For the end-polarized initial state,
\begin{eqnarray}
J_0^{end}(t) &=& \frac{2}{N+1} \sum_k \sin^2(\psi_k)
\cos^2(4bt\cos\psi_k), \nonumber \\
J_2^{end}(t) &=& \frac{1}{N+1} \sum_k \sin^2(\psi_k)
\sin^2(4bt\cos\psi_k).
\label{eq:j02end}
\end{eqnarray}
In both Eqs. (\ref{eq:j02th})-(\ref{eq:j02end}), the normalization is
chosen such that $J_0+2J_2=1$.

\subsection{Experimental results}

\begin{figure} 
\includegraphics[width=3.25in]{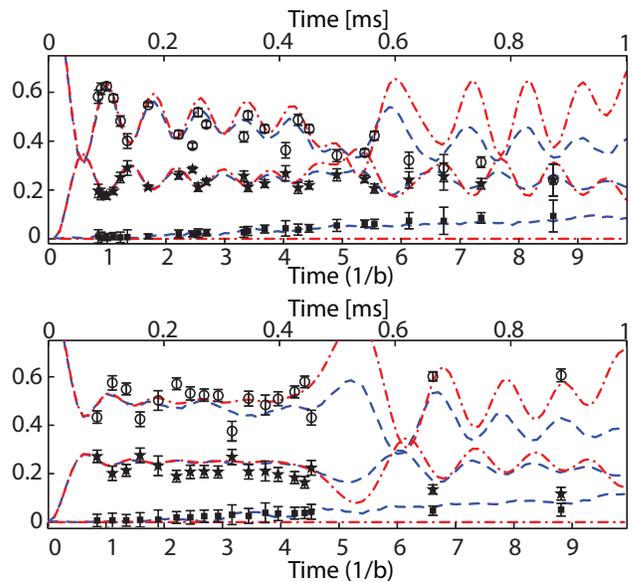} 
\caption{(Color online) Experimental data (0Q circles, 2Q-stars, 4Q
squares), analytical results (red dash-dotted lines), and least square
fit (blue dashed lines) of experimental data to numerical calculations
under the DQ Hamiltonian with NN and NNN couplings of strength $b$ and
$b/8$, respectively.  Error bars for the experimental data were
estimated from the standard deviation of the odd quantum coherences
from zero.  Numerically, two free parameters, the time origin $t_0$
and the time scale $1/b$, are adjusted to obtain the best fit to the
experimental data for $J_0$. Top: Thermal initial state with chain
length $N=18$.  Fitting parameters $t_0= 9.6 \mu$s and $1/b= 101.8
\mu$s. Bottom: End-polarized initial state wth chain length $N=19$.
Fitting parameters $t_0= 0.2 \mu$s and $1/b= 100.8 \mu$s. }
\label{fig2:expdata} 
\end{figure}

The experiments were performed in a 7 T magnetic field using a Bruker
Avance Spectrometer equipped with a home-built probe. The $^{19}$F
frequency is 282.37 MHz.  Experimentally measured MQC data are shown
in Fig.~\ref{fig2:expdata}, along with analytical predictions and
 simulation results under the DQ Hamiltonian with NN and NNN
couplings.  Both the time origin $t_0$ and the coupling strength $b$
were used as fitting parameters in order to minimize the square of the
difference between the experimental and numerical data, that is,
$\sum_i\left|J_0^{\text{exp}}(t_i)-
J_0^{\text{num}}[b(t_i-t_0)]\right|^2$.  For the thermal initial state
data (upper row in Fig. \ref{fig2:expdata}), the $\pi/2$ pulse length
was 1.05 $\mu$s, the inter-pulse delay $\Delta$ was varied from 2.9 to
5 $\mu$s, and the number of loops was increased from 1 to 7.  We set
$K = 12$, and incremented the phase in steps of $2\pi/24$ to encode
the MQCs.  For the end-chain initial state data (lower row in Fig.
\ref{fig2:expdata}), the $\pi/2$ pulse length was 0.93 $\mu$s, the
end-state preparation time $t_1=30.3\: \mu$s, the inter-pulse delay $\Delta$
was varied from 2.9 to 7.3 $\mu$s, and the number of loops increased
from 1 to 8.  We set $K=16$, and incremented the phase in steps of
$2\pi/32$ to encode the MQCs.  In both cases, the recycle delay was
300s and a solid echo sequence with an 8-step phase cycle was used to
read out the signal intensities at the end of the experiment. 

The experimental data are normalized at every time step, such that
$J_0 + 2(J_2 + J_4) = 1$ (using the fact that $J_{-n} = J_n$). The
intensities of the odd MQCs (not shown in Fig.~\ref{fig2:expdata})
turn out to be negligibly small.  At short times (less than $\approx 0.2$
ms), Fig.~\ref{fig2:expdata} indicates that fourth- and higher-order
even-MQCs are also negligible.  However, the four-quantum coherence
signal contributes significantly at longer times.  In 3D systems,
including both plastic crystals such as adamantane \cite{Suter04} and
rigid crystals such as the cubic lattice of $^{19}$F spin in CaF$_2$
\cite{Levy92,Cho05}, very high coherence orders are seen to develop
over a time scale less than a millisecond, with no apparent
restriction on the highest order reached.  In contrast, the fact that
the MQ intensities are restricted to the zero- and DQ-coherences, and
that the higher-order terms only grow relatively slowly during the
whole time domain we explored, are strong indications of the 1D
character of the spin system.  At the same time, the appreciable
intensity of the four-quantum coherence at long evolution times
clearly indicates that the analytical model (which predicts only zero-
and DQ- coherences for both the thermal and end-polarized initial
states) becomes inadequate to accurately describe the real system.

Note that in the simulations, the maximum computationally accessible
chain length was $N=21$ spins.  Though the fits included in Fig. 2 use
$18$-$19$ spins, it is important to realize that sensitivity of the
dynamics to the precise value of $N$ develops only at sufficiently
long times (as the effect of the finite chain boundaries manifest --
see Appendix \ref{sec:mirror}), where the accuracy of the simple model
used to make the estimate becomes itself limited.

\section{Multiple quantum dynamics: Beyond spin-chain approximation}
\label{sec:beyond}

In order to understand the discrepancies observed between the
analytical model and the experimental results, it is necessary to
identify the dominant sources of non-ideality in the experiment, and
assess their respective effects on the observables under examination.
With this in mind, we first analyze effects due to limited control,
such as the higher-order terms in AHT as well as imperfect system
initialization.  We then investigate the intrinsic limitations of the
1D NN model to describe the real physical system, which contains an
ensemble of weakly-coupled spin chains with long-range intra-chain
couplings.  In particular, we compare the effects of long-range
interactions first within a single chain and then across different 
spin chains.  Note that while long-range couplings have been
previously accounted for in a perturbative limit~\cite{Feldman97}, we
resort here to exact numerical simulations (Appendix~\ref{sec:num}),
while also considering other experiment-related sources of errors.

\subsection{Errors due to limited control}
\subsubsection{High-order terms in Average Hamiltonian Theory}

\begin{figure}[b] 
\includegraphics[width=3in]{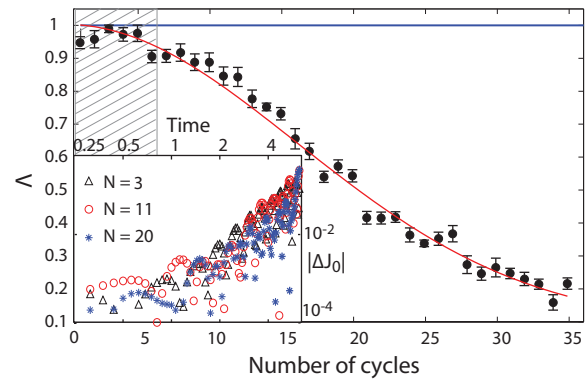}
\caption{(Color online) Overlap $\Lambda$ for end-polarized initial
state, Eq. (\ref{eq:fid}). Normalized experimental data (stars) and
numerical results for finite-width pulses (solid line).  The chain
length in the numerical calculation is $N=9$, whereas the pulse length
and inter-pulse delay are $w = 0.0075\,b\approx 1 \mu$s and
$\Delta=0.0225\,b$, respectively, which are close to the experimental
values.  The shaded area is the time region explored in the MQC
experiments. Inset: Absolute value of the difference between $J_0$ as
calculated from the analytical model or the rf-pulsed dipolar
Hamiltonian with the initial thermal state under different chain
lengths, as a function of cycle time. Only NN couplings are considered
in this case. }
\label{fig3:FidDecay}
\end{figure}

As mentioned, experimentally the DQ Hamiltonian~(\ref{eq:dq}) is
obtained as the zeroth order average Hamiltonian of a multiple-pulse
sequence.  Since the 16-pulse cycle used in the experiment is
time-symmetric~\cite{Ramanathan03}, all odd-order corrections are
zero, and the leading error term is of second-order in $\left\|\bar
{\cal H}^{(2)}T_c\right\|$, both when considering ideal and
finite-width pulses.  The contributions
to the effective Hamiltonian from higher-order terms may be estimated
by comparing the single-cycle MQC signal computed using the exact DQ
Hamiltonian, and using the dipolar Hamiltonian (\ref{eq:dip})
interspersed with rf pulses, respectively. Assuming ideal
instantaneous pulses, we verified numerically that for the system of
interest such contributions are small provided that the cycle time
$T_c \lesssim 4$ (see Fig. \ref{fig3:FidDecay}, Inset).  In the
experiment, we thus employed multi-cycle sequences, in order to extend
the region of validity of the DQ model.

In order to determine how well we implemented the evolution reversal
experiment described in Sec. II, we performed a series of experiments
that measured the overlap between the initial and the final state,
following evolution reversal.  This overlap is given by
\begin{equation}
\Lambda = \tr[\rho_{th} U_{MQ}^y U_{MQ}^x \rho_{end}(0)
U_{MQ}^{x\dag} U_{MQ}^{y\dag}], \label{eq:fid}
\end{equation}
where $\rho_{end}(0)$ is the end-polarized state, and the observable
is the collective magnetization $\rho_{th} = \sigma_z$. To lowest
order, $U_{MQ}^y$ (see Eq. (\ref{MQpropagator})) is approximately the
inverse of $U_{MQ}^x$. Thus, the overlap $\Lambda$ is close to maximal
for short cycle times. The experimental data is shown in
Fig.~\ref{fig3:FidDecay}.  The experimental data were normalized by fitting
the decay to a normalized Gaussian curve.  
The $\pi/2$ pulse length used was $w=0.93\;\mu$s, whereas the
delay $\Delta=2.9\;\mu$s.  In normalized units (the NN coupling $b
\approx 8.3$~kHz in practice), this corresponds to $T_c \approx 0.72$,
indicating that we are well within the regime where the contributions
of the higher order terms can be neglected.  Even as $\Delta$ is
increased to 7.3 $\mu$s in some of the experiments, $T_c$ only
increases to $\approx 1.64$ (in normalized units), thus still within
the range where higher-order corrections are unimportant.

This is confirmed by numerical simulations, also shown in the main
panel of Fig.~\ref{fig3:FidDecay}. We prepared the end-polarized
initial state in a matrix form for a system of 9 spins and evolved the
system first forward under the DQ sequence with pulses along the
$x$-axis, then backward by using $y$-pulses.  Considering that in
practice the DQ coupling strength $b_{ij}\lesssim 8.3$ kHz, and that
finite-width corrections originate primarily from the second-order
average Hamiltonian, we expect these corrections to be on the order of
$(b_{ij}w)^2 \lesssim 6.9\times 10^{-5}$.  As seen in
Fig.~\ref{fig3:FidDecay}, the overlap from numerical calculations is
flat and close to unity, confirming that errors due to finite widths
and high-orders AHT contributions are small.  Comparison with the
experimental data suggests that other sources of error are likely to
be responsible for the long-term decay of the overlap
\cite{Haeberlen76}.  In particular, both rf and static-field
inhomogeneities can result in imperfect $\pi/2$-pulses, leading to
off-axis and pulse-length systematic errors.  The latter errors are
actually minimized by the 16-pulse sequence thanks to the use of phase
alternation~\cite{Slichter92}. Furthermore, transient effects of
square pulse always exists in pulse-driven experiments.  Notice that
the MQC data of Fig.~\ref{fig2:expdata} were measured at relatively
short times, $t\lesssim 0.5$ ms for most of the data.  This
corresponds to 6 cycles, thereby to high values of the overlap.

\subsubsection{Initialization}
\label{sec:endprep}

\begin{figure}
\includegraphics[width=3.2in]{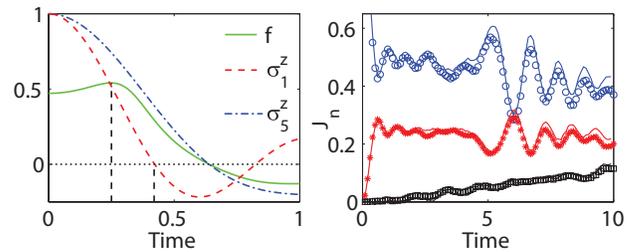}
\caption{(Color online) Left panel: Evolution of the end- and central-
spin polarizations and fidelity during the preparation of the
end-polarized state. Chain length $N=9$, $\alpha = x, \bar y$, and
$N_\alpha=2$. The two time points $t_1=0.25$ and $t_1'=0.42$ are
marked with vertical dashed lines. Right panel: MQC signal of $N=19$
spin chain with preparation time $t_1=0.25$. Intensities are
normalized at every time as $J_0+2(J_2+J_4)=1$. DQ Hamiltonian with
NN+NNN couplings and the ideal end-polarized initial state -- solid
lines; DQ Hamiltonian with NN+NNN couplings and initial state
synthesized at $t_1=0.25$ -- circles for $J_0$, stars for $J_2$, and
squares for $J_4$.}
\label{fig4:EndChainPrep}
\end{figure}

The basic idea for preparing the end-polarized initial state from the
thermal state was introduced in~\cite{Cappellaro07a}. Starting from
equilibrium, we first rotate the nuclear spins into the $x$-$y$ plane
by a $\pi/2$ pulse along a direction $\alpha$. We then allow the
system to evolve under the dipolar Hamiltonian of Eq.~(\ref{eq:dip})
for a time $t_1$ ($=30.3$ $\mu$s in the experiment, corresponding to
0.25 in normalized units), and finally rotate the spins back to the
$z$-axis by a second $\pi/2$ pulse along the $-\alpha$
direction. During time $t_1$, the spins at both ends evolve roughly
$1/\sqrt 2$ times slower than the internal spins, due to the fact that
each of them has only one nearest neighbor, while any internal spin
has two.  Let $U_\alpha$ describe evolution under the pulse sequence
${\pi}/{2}\left|_{\alpha} - t_1 - {\pi}/{2}\right|_{\bar \alpha}$,
where in the experiment the pulse axis $\alpha$ is phase-cycled
through the $y$- and $x$-axes.  Given that the state at time $t_1$ is
$\rho(t_1) = (1/N_\alpha) \sum_\alpha U_\alpha\rho_{th}
U_\alpha^\dag$, with $N_\alpha$ being the number of phase-cycling
steps, the fidelity of the prepared $\rho(t_1)$ relative to the
desired end-polarized state is
\begin{equation}
f(t_1) = \frac{\tr[\rho_{end} \rho(t_1)]}{\sqrt{\tr[\rho_{end}^2]
\tr[\rho^2(t_1)]}}\;.
\end{equation}
The difference between $\rho(t_1)$ and $\rho_{end}$ is due to the
presence of zero quantum coherences which are generated by the dipolar
Hamiltonian but are not be removed by phase-cycling, with leading
contributions from residual polarization on spins $2$ and $N-1$, as
well as correlated states of the form $\sigma_i^z(\sigma^+_{i-1}
\sigma^-_{i+1}+\sigma^-_{i-1}\sigma^+_{i+1})$~\cite{Cappellaro07a}.
The left panel of Fig.~\ref{fig4:EndChainPrep} depicts the time
dependence of the fidelity and the polarization of the end and the
central spins.  Interestingly, the time that maximizes fidelity
($t_1=0.25$) does {\em not} coincide with the time at which the
central-spin polarization is zero ($t_1'=0.42$). It is also worth
mentioning that both time points are almost independent of the chain
length unless $N \leq 4$.

Starting from the two prepared states, $t_1=0.25$ and $t_1'=0.42$,
respectively, we calculate the MQC of the spin chain under the DQ
Hamiltonian with NN+NNN couplings, and compare the results against those
obtained for the ideal end-polarized initial state $\rho_{end}(0)$. The
evolution of MQC for the initial state prepared with $t_1'=0.42$ is
quite different from that obtained with the intended state (data not
shown), while the MQC of the initial state corresponding to
preparation time $t_1=0.25$ is very close, as demonstrated in the
right panel of Fig.~\ref{fig4:EndChainPrep}. Note, however, that
compared to the ideal end-polarized state, the experimentally prepared
initial state shows slightly larger oscillations, especially in $J_0$.

\begin{figure*}[ht] 
\includegraphics[scale=.6]{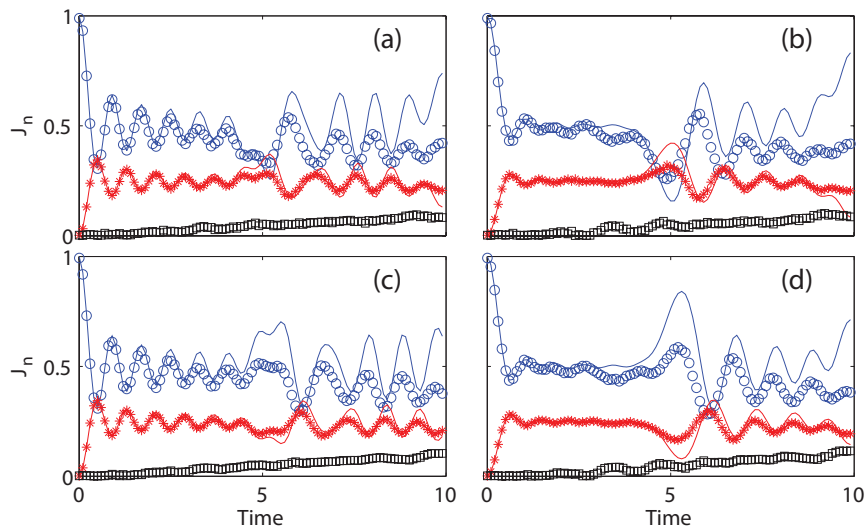}
\caption{(Color online) Effect of NNN interactions in the DQ
Hamiltonian for the thermal (left column -- a and c) and the
end-polarized initial state (right column -- b and d). The length of
the spin chain is $N=18$ (top row -- a and b) and $19$ (bottom row --
c and d). Solid (blue and red lines) are $J_0$ and $J_2$ computed from
the analytical model, respectively. Circles, stars, and squares are
$J_0$, $J_2$, and $J_4$ obtained from exact numerical results.}
\label{fig5:nnn}
\end{figure*}

\subsection{Non-idealities in isolated single-chain dynamics}
\subsubsection{Long-range couplings}
\label{sub:long-range}

As already remarked, appreciable growth of the four-quantum coherence
signal at long times (Fig.~\ref{fig2:expdata}) indicates the
inadequacy of the analytical model, as the dynamics is no longer
confined to a 1D system with pure NN couplings.  Having shown in the
previous section that the effect of higher-order AHT terms
is negligible in the temporal region we consider, we next turn our
attention to the influence of long-range couplings. We limit our
calculations to the NNN couplings as they provide the most important
correction to the analytical NN model.

Figure~\ref{fig5:nnn} shows the MQC signal obtained for spin chains of
length $N=18$ (top) and $19$ (bottom) for the thermal state (left) and
the end-polarized state (right), respectively. Both NN and NNN
couplings in the DQ Hamiltonian are now exactly accounted for.  By way
of comparison, we also include the predictions from the analytical
model. The following main observations may be made:

(i) NNN couplings produce even-order coherences greater than two, the
largest contributions in the relevant time window arising from
$J_4$. In general, even order coherences up to the number $N$ of spins
in the chain may be expected. This is in contrast with the results
based on a perturbative approach~\cite{Feldman97}, which yield MQC
only up to the sixth order.

(ii) NNN couplings reduce the amplitude of the oscillations in $J_0$
and $J_2$.

(iii) The effect of NNN couplings is amplified at an instant in time
that we call the {\em mirror time}, $t_m$ ($\approx 5$ in the figure),
which is defined in terms of the analytical model as the time where
$J_0$ shows a second largest oscillation for odd $N$ or the lowest
point for even $N$. (Note that one could also equivalently define
$t_m$ as the time where the second lowest/largest peak of $J_2$
occurs.) This effect is prominent in the numerical simulations, where
one is necessarily constrained to relatively short chains.
Qualitatively (see also Appendix B), the spin dynamics has a mirror
symmetry about the middle spin, which causes the signal of specularly
located spins to ``interfere constructively'' at the mirror time.
This picture can also explain why the influence of NNN couplings on
the dynamics of the chosen collective observable is most pronounced at
this time: even small deviations from the ideal NN dynamics are able
to destroy the interferences and can produce significant changes in
the observed signal.

\subsubsection{Chain length distribution}

\begin{figure}[t]
\includegraphics[width=3.25in]{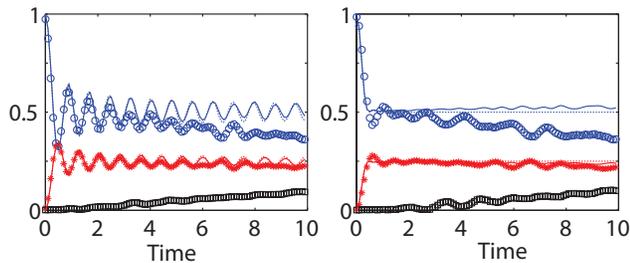}
\caption{(Color online) Ensemble average of MQC signal for the thermal
(left) and the end-polarized initial state (right), as resulting from
a DQ Hamiltonian with NN and NNN couplings.  Circles, stars, and
squares show $J_0$, $J_2$ and $J_4$, respectively, with chain length
averaged over $N=[17,21]$. For comparison, the solid lines are for a
distribution of chains with average $\bar{N}=50$ in the random cluster
model, evolved under the analytical model, whereas the dotted lines
are for the zero and double quantum intensities for an infinite
chain~\cite{Feldman97}.}
\label{fig6:ensemble}
\end{figure}

Since the defects in the FAp sample are non-uniform, the spin chain
length has a statistical distribution. According to the so-called
random cluster model~\cite{Cho96}, if defects are distributed randomly
in the infinite 1D chain with a probability $(1-p)$, the average chain
length is $\bar N = (1+p)/(1-p)$, and the relative fluctuation $\Delta
N / \bar N = \sqrt{2p} \;/ (1+p)$.  For a low percentage of defects,
($p\approx 1$, $\bar N\gg 1$), the chain length distribution can be
reasonably approximated by a uniform distribution of chain lengths.
Fig.~\ref{fig6:ensemble} shows the averaged MQC signal for an ensemble
of chain lengths.  Compared to an individual spin chain, the ensemble
average washes out the long-time oscillations but leaves the
short-time oscillations virtually unchanged. Since the concentration
of defects is low in the actual sample, we expect that this effect
will not be important on the time scales explored by the current
experiments.

\subsection{Non-idealities due to coupled-chain dynamics}

Due to the 3D nature of the FAp sample, a given spin chain of interest
(``central'' spin chain henceforth) is coupled to all other chains in
the crystal via the long-range dipolar coupling.  Since the distance
between two spin chains in FAp is about three times the distance of
two NN $^{19}$F spins, the {\em cross-chain} couplings have about the
same strength as the third-neighbor intra-chain coupling within a
chain. The combined effect is, however, amplified by the presence of
several (six) chains surrounding the central spin chain (recall
Fig. 1).  Furthermore, additional weaker contributions arise from more
distant chains. Overall, the influence of the cross-chain coupling can
thus be an important source of deviation from the analytical model, as
we explore next.

Exactly modeling the influence of all chains on the central one would
require us to simulate the quantum dynamics of a macroscopically large
number of spins, which is clearly beyond reach.  To make the problem
numerically tractable, we thus need to reduce the many-body problem to
a simpler model that represents as faithfully as possible those
features of the real dynamics we are directly probing.  In order to
make sensible approximations, it is useful to reconsider the origin of
the NMR signal in more detail.  Let $M_c$ be the number of chains
present in the crystal sample. In the high-temperature approximation,
the initial density matrix of the whole system can be expressed as
$$ \rho_{\text{3D}} (0) = \sum_{m=1}^{M_c} \rho_m(0),$$
\noindent 
where $m$ indexes the chains and $\rho_m(0)$ is either the thermal
equilibrium state or the end-polarized state, as in
Eqs. (\ref{eq:ts})-(\ref{eq:endp}).  Notice that due to its collective
nature, the experimentally accessible observable can also be written
as a sum of contributions from distinct chains.  For the purpose of
making contact with a reduced description where a single chain is
singled out as a reference in the presence of other coupled chains, it
is useful to view the total signal in the $k$-th measurement $S_z^k$
as originating from two terms, $S_z^k (t) = S_{z,intra}^k +
S_{z,leak}^k$, with
\begin{subequations}
\begin{equation} 
S_{z,intra}^k (t) =\sum_{m=1}^{M_c} {\rm Tr} \Big[\sum_j
\sigma_{mj}^z\rho_{m}^k(t)\Big], 
\label{eq:intra}
\end{equation}
\begin{equation} 
S_{z,leak}^k (t) = \sum_{m'\neq m}{\rm Tr} \Big[\sum_j \sigma_{m'j}^z
\rho_{m}^k(t)\Big],
\label{eq:leak}
\end{equation}
\end{subequations}
where $\rho^k_{m}(t) = U_k(t) \rho_m(0) U_k^\dag(t)$ (cf.
Eq.~(\ref{eq:Jn})).  These two terms reflect two different mechanisms
by which the presence of cross-chain interactions can induce
deviations of the experimental signal from that of an isolated chain.

The term in Eq. (\ref{eq:intra}), which we refer to as the
\textit{intra-chain signal} $S_{z,intra}^k (t)$, describes the signal
obtained when the initial state and observable belong to the same
chain: all other chains, which may initially be taken to be in the
maximally mixed state, influence the reference chain in a ``mean-field
sense,'' to the extent they modify $\rho^k_{m}(t)$.  Were all the
chains identical, the resulting signal would simply be $S_{z,intra}^k
(t) \approx M_c {\rm Tr} [ S_z \rho^k (t)]$, that is, an $M_c$-fold
signal from a single chain coupled to the ``environment chains''.
Thus, $S_{z,intra}^k (t)$ may be well described within a
``chain-plus-environment'' model, where a single spin chain is coupled
to a larger spin environment, and the measured NMR signal is
determined completely by the reduced density matrix of the central
chain -- upon tracing over all environment spins, as in a standard
formulation of the central-system-plus-bath problem
\cite{CentrPlusBath,FeynmanVernon,SixManPaper}.

While the intra-chain term describes a deviation from the ideal
single-chain behavior that is not fundamentally different from, say,
deviations induced by long-range couplings as analyzed in
Sec. \ref{sub:long-range}, the \textit{leakage signal} $S_{z,leak}^k
(t)$ in Eq.~(\ref{eq:leak}) introduces a qualitatively different
effect: that is, the possibility that some of the polarization
initially located on the $m$th chain is transferred to the $m'$th
chains, and read out there.  Since, from the point of view of the
central spin chain, signal components would be `lost' to the
environment, a significant contribution $S_{z,leak}^k (t)$ would
clearly indicate the inadequacy of a system-plus-environment picture
at capturing the complexity of the underlying 3D strongly-correlated
dynamics.

Even assuming that the consistency of a central chain-plus-environment
treatment may be justified {\em a posteriori} by the smallness of the
leakage signal for the evolution times of interest, modeling a
realistic environment remains non-trivial because the actual crystal
consists of a large number of quantum spin chains, evolving according
to a highly complex, non-Markovian dynamics.  In line with standard
statistical approaches (including NMR relaxation theories)
\cite{Kubo,Slichter92}, we can however reasonably argue that the main
observed features should be robust with respect to the details of the
environment description, as long as the relevant energy scales are
correctly reproduced.  In what follows, we will exemplify these
considerations by separately investigating two models for describing
how the coupling between different chains in FAp modifies the MQC
dynamics of the central spin chain.  In Sec.~\ref{sec:TwoChains}, a
system of two coupled chains is investigated as a numerically
accessible testbed where the `environment chain' qualitatively retains
the spatial structure of the FAp crystal.  Physically, the latter
feature is expected to be important (possibly essential) to properly
represent the deviation induced in the idealized central-chain
dynamics by the NN chains.  In Sec.~\ref{sec:ChainPlusBath}, a
structureless spin environment model is considered instead, whereby
the central chain couples to randomly placed spins.  Physically, such
a picture may be especially adequate to account for the net influence
of distant chains. Computational constraints limit the size of the
accessible model environment in both cases.

In spite of the above differences, it is important to realize that
essentially the same type of simulations will be employed and the same
main physics will be explored in both cases.  In particular, the
processes leading to deviations from the analytical model are
primarily associated with the increased dimensionality of the Hilbert
space and correlations between different chains in the sample.  While
{\em no} explicitly non-unitary evolution is present either in
experiment or simulation, and the total system remains coherent at all
times, a damping of the low-order MQC oscillations still emerges: as
time progresses, a larger part of the Hilbert space is populated, and
coherences of higher-order, which involve spins of different chains,
build up at the expenses of low-order coherences.  Relative to the
observables that can be directly probed, the latter simply appear to
unrecoverably decay.

\subsubsection{Effect of a structured spin environment}
\label{sec:TwoChains}

The correlated dynamics in nearby chains may be investigated by
lumping together the contributions of the six nearest surrounding
chains and treating them as a single chain, which couples coherently to
the central spin chain, according to the DQ Hamiltonian. We take both
chains to have length $N$ and start in the initial state of interest
(either thermal or end-polarized).  Since we are restricted to
numerically calculate MQC for a system of up to $25$ spins, $N\lesssim
12$ in practice.  Upon summation [$M_c=2$ in
Eq. (\ref{eq:intra})-(\ref{eq:leak})], an {\em upper bound} to the NN
cross-chain coupling strength is given by
\begin{eqnarray} 
\frac{\bar b_\times}{b} = \frac{\sum
b_\times}{b} &=& -3 \left( \frac{d}{D}\right)^3 \approx -0.1488.
\label{eq:bcross} 
\end{eqnarray}
\noindent
This approximation corresponds to neglecting correlations between
spins from three or more different chains, which arise from
higher-order cross-chain couplings in $S_{z,leak}^k (t)$ -- for
instance, the three-chain coupling is proportional to $(\bar{b}_\times
/ b)^2$.  If such couplings are treated perturbatively, one may expect
their effect to be negligible over the time scale of the experiment,
as opposed to two-chain interactions which directly compete in
strength with intra-chain NNN couplings.  As discussed above, however,
these two contributions may have very different physical implications,
as the cross-chain coupling effect can genuinely increase the
underlying Hilbert space, whereas NNN couplings can only increase the
portion of the single spin-chain Hilbert space that is explored during
the dynamics.  

\begin{figure}[t] 
\includegraphics[width=2.9in]{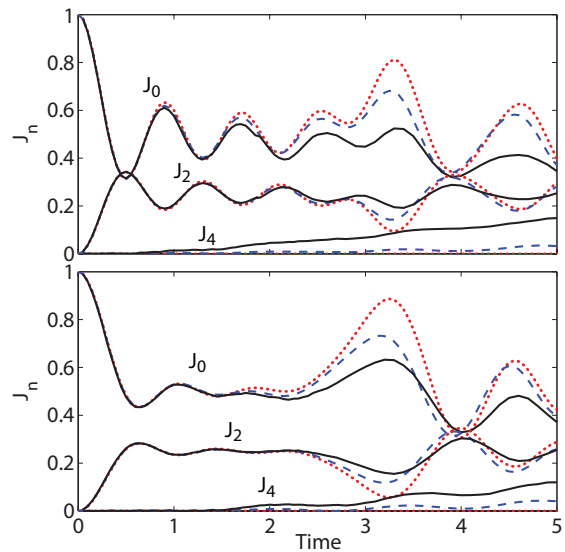} 
\caption{(Color online) Effect of cross-chain couplings for the
thermal (top) and the end-polarized state (bottom). The length of each
spin chain is $N=11$. Dotted lines represent the analytical
isolated-chain NN prediction. Both NN and NNN couplings are considered
in the exact simulation results for a single chain (dashed lines) as
well as for two chains (solid lines) coupled according to
Eq. (\ref{eq:bcross}).}
\label{fig7:twochains} 
\end{figure}

Exact calculation of the total signal $S_{z}^k (t)$ reveals that the
contribution of cross-chain transfer due to $S_{z,leak}^k (t)$ remains
small  (below a few percents) over relatively short time scales
(up to 5 in normalized units).  As shown in Fig.~\ref{fig7:twochains},
cross-chain couplings modeled in this way also damp the MQC
oscillations at long times, similar to the effect of intra-chain NNN
interations.  Notice that at the mirror time, as observed in
simulations with finite $N$, the effects of the cross-chain couplings
are also amplified, further reducing the peak amplitude
\cite{comment}.

\subsubsection{Effect of a structureless spin environment}
\label{sec:ChainPlusBath}

According to our earlier discussion, another way to analyze the
influence of cross-chain coupling that emphasizes the influence of far
away chains is to consider an effectively structureless quantum spin
environment \cite{BathRemark}.  In particular, the simplest choice is
provided by a system consisting of spins arbitrarily (randomly)
scattered in space.  While of course nothing is arbitrary in the
dynamics of the real FAp system, such a randomized model may just be
viewed as a computationally accessible approximation of the complex
dynamics under investigation.

\begin{figure}[tb]
\includegraphics[scale=0.2,angle=-0]{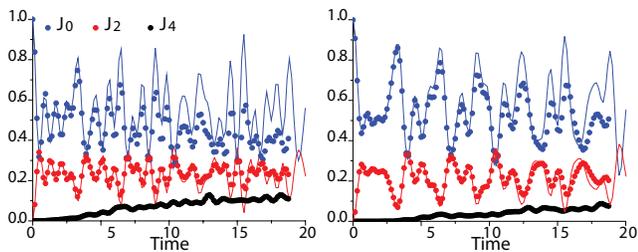}
\caption{(Color online) $J_0(t)$, $J_2(t)$, and $J_4(t)$ for a chain
of $N=11$ spins with thermal (left) and end-polarized (right) initial
state.  The chain is coupled to an environment of nine spins with
random dipolar couplings.  The lines are the analytical results for
$J_0(t)$, $J_2(t)$.  Note that only NN intra-chain couplings are
included, however the times scales explored here are significantly
longer than in any of the previous figures.}
\label{fig8:randombath}
\end{figure}

\begin{figure}[t] 
\includegraphics[width=2.9in]{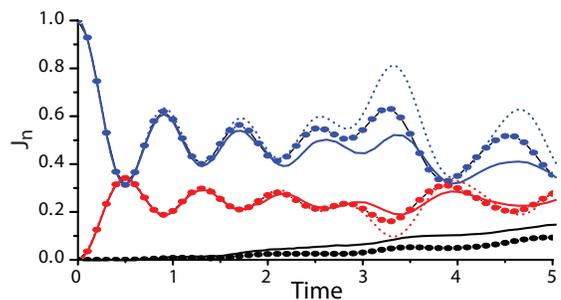} 
\caption{(Color online) Comparison between the effect of cross-chain
couplings starting from the thermal state, as resulting from a
structured two-chain system (solid lines) or from a coupling to a
randomized spin (circles). In all cases, the length of the spin
chain(s) is $N=11$ and NNN intra-chain couplings are included. Dotted
lines represent the analytical prediction.  Numerical results for
$J_0(t)$ (blue), $J_2(t)$ (red), and $J_4(t)$ (black) MQCs are
included.}
\label{fig9:twoch_randbath} 
\end{figure}

Specifically, we reproduce the main features and the characteristic
energy scales of the FAp sample driven by the $16$-pulse sequence by
assuming that the $x$, $y$, $z$ coordinates of each of the 9
environment spins are drawn uniformly from $[-1,1]$.  The $N=11$ spins
of the central chain are placed equidistantly on the $z$-axis, with
their $z$-coordinates also confined between $-1$ and $1$.  The minimum
distance between any pair of spins (whether environment or chain
spins) is restricted to exceed 0.1 to prevent spins from being too
close to each other.  The central chain Hamiltonian of the form
(\ref{eq:dip}) is truncated at either the NN or NNN level
(Fig. \ref{fig8:randombath} and \ref{fig9:twoch_randbath}
respectively).  All the dipolar coupling between the environment
spins, and from the environment spins to the central spin chain are
taken into account, as in Eq.  (\ref{eq:dip}), with the coupling
constants $b_{j\ell}$ calculated from the spins coordinates.  However,
in order to have correct energy scales, all chain-environment coupling
constants are rescaled to produce the correct value of the dispersion
${\rm Tr}\,[H^2_{CB}]$, where $H_{CB}$ is the chain-environment
interaction Hamiltonian (see also Appendix D).  This ensures that the
couplings between the spins of different chains in FAp are $\approx
40$ times smaller than the couplings between the spins in the same
chain. In a similar way, all couplings inside the environment are
rescaled to produce a correct value for the Hamiltonian norms per
spin, ${\rm Tr}\,[H^2_C]={\rm Tr}\,[H^2_B]$, where $H_C$, $H_B$ are
the chain and environment Hamiltonians, respectively.

We perform simulations of the total system treating it as a closed
quantum system with unitary dynamics (see Appendix \ref{sec:num} for
details).  We simulate the evolution under the experimental DQ
Hamiltonian, generated by the 16-pulse sequence. Each sequence with 16
pulses along the $x$-axis (preparation) and, afterward, the 16-pulse
sequence with pulses along the $y$-axis is repeated five times
(mixing).  The pulses are ideal $\delta$-like, with varying
inter-pulse separation, and the total time is varied from zero to
18.75 (in normalized units).  Note that since the environment is
homonuclear, it is affected by the pulses in the same way as the
central chain.  At the end of each protocol, the total NMR signal
$S_{z}^k (t)$ is calculated by either (i) summing only the
$z$-projections of the spins in the central chain, tracing out the
environment spins (thus obtaining only the intra-chain contribution of
Eq. (\ref{eq:intra}); or (ii) summing the $z$-projections of both the
environment and the chain spins (thus also taking into account the
leakage terms in Eq. (\ref{eq:leak})).  Comparison between the results
(i) and (ii) shows that the leakage terms are small, on the order of
about 1\%.  As in the two-chain model, we thus confirm {\em a
posteriori} the validity of the underlying {\em weak-coupling}
assumption between the central system and the rest.

Numerical results starting from the thermal and the end-polarized
state are given in Fig.~\ref{fig8:randombath} for a {\em single}
realization of such a random dipolarly-coupled environment,
corresponding to a fixed (arbitrary) geometry of the spin lattice.
While different realizations give very close results (data not shown),
averaging over several realizations is impractical \cite{note}.
Fig.~\ref{fig8:randombath} also includes a comparison of the
simulation results for $J_{0,2}(t)$ with the analytical model.  The
interaction with the environment leads to a significant damping of the
oscillations of $J_0(t)$ and $J_2(t)$, and to an overall decay of
these coherences.  Interestingly, the decay of both $J_0(t)$ and
$J_2(t)$ for the end-polarized initial state is {\em slower} than for
the thermal initial state.  Likewise, it is also worth noticing that
the decay of the oscillations in $J_2$ is roughly a factor of two
slower than the decay of $J_0$. This difference may be attributed to
the fact that the random dipolarly-coupled environment is {\em not
fully structureless}, as it possesses non-trivial integrals of motion
(for instance the total magnetization of the central chain and the
environment). The internal structure of the environment appears to
strongly affect the dynamics of $J_2(t)$. This different behavior of
the two MQC intensities is also present in the two coupled-chain
simulations of Sec. \ref{sec:TwoChains}, which are directly contrasted
to the random spin-environment simulation results in
Fig. \ref{fig9:twoch_randbath}.  We further expand on these
considerations by examining a chaotic spin bath model in Appendix
\ref{sec:ChBath}.

\section{Discussion and conclusion}
\label{sec:con}

We have investigated in detail the MQC dynamics of a quasi-1D spin
chain in a fluorapatite crystal, both experimentally and
numerically. By comparing exact simulation results with analytical
solutions for the ideal DQ Hamiltonian with NN couplings, we have
characterized the region of validity of this simple, single-chain NN
model.  For the initial states and observables of interest, we have
found that for evolution times up to 0.5 ms (corresponding to about 5
times the inverse NN coupling strength) the system is experimentally
indistinguishable from the single-chain, NN model.  Simulations
including long-range couplings within a single chain and across
different chains reproduce well the experimental findings.
 
Beyond this time, the evolution deviates from the analytical model,
although the deviations of the selected observables (the MQC) remain
small.  In principle, the experimental implementation of the DQ
Hamiltonian using a simulation approach based on AHT is not a problem,
at the evolution times considered.  In addition, the dynamics of the
experimentally created end-polarized initial state are seen to remain
quite close to the dynamics of an ideal end-polarized state, as
desired.

From simulations we observed that all the different types of
long-range couplings analyzed lead to a qualitatively similar damping
of the oscillations in the MQC signal and a relatively slow growth of
the higher order coherences (in particular the 4-quantum coherence).
In fact, a similar effect is also observed for a single chain coupled
to a dipolar spin environment.

The similar behavior observed when introducing longer range couplings
in a 1D chain and cross-chain couplings seems to indicate that
although in the second case there are more pathways available for the
propagation of multi-spin correlations, this effect cannot be observed
in the MQC evolution.  While it could be tempting to infer that the
microscopic mechanisms leading to the observed behavior are to some
extent similar in each case, it is also essential to acknowledge that
the experimentally accessible, collective magnetization observable
provides a highly {\em coarse-grained} visualization of the overall
dynamics.

From a many-body physics standpoint, a deeper understanding of the
influence of the structure of the longer-range dipolar couplings
(``internal environment'') on MQC dynamics, in particular of the
potentially higher level of sensitivity found for higher-coherence
orders, is certainly very desirable.

Lastly, from a quantum communication perspective, our work calls
attention to the added challenges that transport protocols need to
face in the presence of limitations in available control,
initialization, and readout capabilities, as well as long-range
interactions and/or unwanted interactions with uncontrolled degrees of
freedom.  Our study points out that for the realization of precise
transport, simply isolating a 1D system is not enough, as the
deviation from an ideal NN model in a 1D chain caused by long-range
couplings is as important as cross-chain couplings.  Since a number of
these issues are shared by all practical device technologies to a
greater or lesser extent, it is our hope that our analysis will prompt
further theoretical investigations of communication protocols under
realistic operational and physical constraints.

\acknowledgments
W.Z., V.V.D., and L.V. gratefully acknowledge partial support from the
Department of Energy -- Basic Energy Sciences under Contract
No. DE-AC02-07CH11358.  Part of the calculations used resources of the
National Energy Research Scientific Computing Center, which is
supported by the Office of Science of the U. S. Department of Energy
under Contract No.  DE-AC02-05CH11231. L.V. is grateful to the {\em
Center for Extreme Quantum Information Theory} at MIT for hospitality
and partial support during the early stages of this work.  P.C. is
funded by the  NSF through a grant to the Institute for 
Theoretical Atomic, Molecular and Optical Physics (ITAMP).  This work was
supported in part by the National Security Agency under Army Research
Office contract number W911NF-05-1-0469.  N.A. and B.P. acknowledge
support from the MIT Undergraduate Research Opportunities Program.

\appendix
\section{Numerical methods}
\label{sec:num}

We calibrate our numerical procedure by reproducing the results from
the analytical model~\cite{Cappellaro07a,Cappellaro07b}. For
sufficiently short spin chains, $N\leq 11$, we propagate exactly the
density matrix of the system. That is, given an initial mixed state
(either the thermal or the end-polarized state), we prepare the
initial density matrix, evolve the system, obtain the density matrix
at time $t$, and calculate the MQC signal according to
Eq.~(\ref{eq:Jn}). For longer spin chains, this approach becomes very
inefficient due to the extremely high usage of computer memory (on the
order of $4^N$ for a chain with $N$ spins). Instead, we employ a
wave-function-based simulation method, whose memory usage scales as
$2^N$. 

To implement the wave-function simulation, we decompose the initial
mixed density matrix of the system into a sum of $N$ (for the thermal
state) or $2$ (for the end-polarized state) individual density
matrices, and then approximate the $j$th density matrix with a product
state of a known state of spin $j$ and a pure random state of the
remaining spins~\cite{Zhang07r,Zhang08}.  That is, we let
$$\sigma_j^z = \frac{1}{2}\Big( {\openone}_j - |\downarrow_j \rangle
\langle \downarrow_j | \Big) \otimes |r_j\rangle\langle r_j|,$$
\noindent
where $|r_j\rangle = \sum_{i=1}^{2^{N-1}} c_i |i\rangle$ is a linear
combination of basis states of all spins except the $j$th spin, and
$c_i$ is a random complex number obeying $\sum_{i=1}^{2^{N-1}} |c_i|^2 =
1$.  Such a superposition is an exponentially
accurate representation of the maximally mixed state,
and in our simulations creates errors of about 0.5\%. 
After preparing the initial wave-function, we propagate the system
according to the Schr\"odinger equation, adopting an efficient
algorithm based on Chebyshev polynomial expansion of the evolution
operator~\cite{Dobrovitski03}.

In the calculation of the MQC signal for the system-plus-environment,
an alternative way to prepare the initial state is used, by realizing
that the initial density matrix may be expressed in terms of spin
operators as follows. Let $|R\rangle = \sum_{i=1}^{2^N} c_i |i\rangle$
be a random wave-function of $N$ spins, and $|R'\rangle = \sum_{j=1}^N
\sigma_j^z |R\rangle$. Then we may simply write $\rho(0) = |R'\rangle
\langle R|$. The propagation of these two wave-functions is then
implemented based on the methods mentioned above.

\section{Mirror Time}
\label{sec:mirror}

\begin{figure}[hbt] 
\includegraphics[width=3.1in]{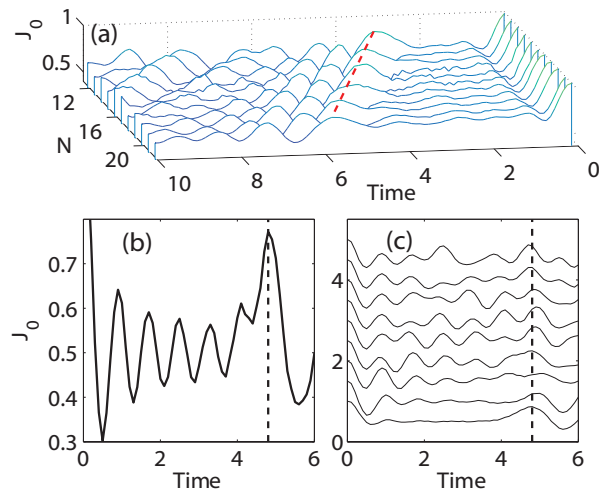}   
\caption{(Color online) (a) $J_0$ for the end-polarized initial state
of different length spin chains (DQ Hamiltonian with NN+NNN
couplings).  The dashed line is a visual guide to the mirror time. (b)
$J_0$ for the thermal initial state and (c) its partition for
different locally polarized initial state $\sigma_i^z$
($i=1,2,\cdots,9$). Due to the mirror symmetry of the chain, only half
of the initial states are presented in panel (c). Each curve is
shifted upward $0.5$ for a better view.  The vertical dashed line
specifies the position of mirror time $t_m$. For (b) and (c), the
length of the spin chain is $N=17$.}
\label{figA1:mirror}
\end{figure}

Besides the peak (dip) of the MQC signal and the amplification of the
NNN coupling effects at the mirror time $t_m$, the following features
may be interesting for spin transport in short spin chains:

(i) The mirror time increases linearly with the length of the spin
chain $N$, as shown in Fig.~\ref{figA1:mirror}(a).

(ii) For the same length spin chain, different locally polarized
initial states have the same mirror time (of course, the thermal
state, which may be seen as a mixture of different locally polarized
initial states, also exhibits the same mirror time); See
Fig.~\ref{figA1:mirror}(b)-(c).

(iii) The NNN couplings shift the mirror time slightly.

These peculiar properties demand a better understanding of the
physical meaning of the mirror time. In a picture of spin polarization
transport along a chain~\cite{Cappellaro07b}, starting from the
end-polarized state where the polarization is pinned to spins $1$ and
$N$, the polarization is transported to the central spin $(N-1)/2$ at
the mirror time $t_m$ (we assume $N$ is odd for simplicity). As
mentioned in the main text, the spin dynamics exhibits a mirror
symmetry about the central spin, and thus interferes constructively at
$t_m$. For other pairs of locally polarized initial states, for
instance $\sigma_j^z$ and $\sigma_{N+1-j}^z$, the spin polarization
also interferes constructively at the mirror time. The independence of
$t_m$ on $j$ guarantees that the thermal state shows the same
properties at $t_m$ as the end-polarized state.

\section{Chaotic Bath Model}
\label{sec:ChBath}

Since in simulations we cannot exactly reproduce the many-body
dynamics occurring in the FAp crystal, approximations are necessary at
a number of levels. In representing the dynamics in terms of a single
chain coupled to a bath the random dipolarly-coupled environment model
used in the main text (Sec.~\ref{sec:ChainPlusBath}) imposes a
structure on the environment that is motivated by the physical system
itself. From an open-system perspective, however, it may be
interesting to explore alternative models for the bath, in order to
have a sense of which details are important for the system's dynamics
and which are not. Although these alternative bath models need not
have an immediate relevance to the experimental system, they may
provide additional physical insight on the action of a spin bath in
the FAp crystal.  In this venue, it is useful to observe that quantum
systems possessing a very complex behavior often exhibit similar
features, and relevant aspects of their dynamics may be captured by
quantum chaotic models, see {\em e.g.} \cite{Guhr98}.  Following this
approach, we emulate the bath's internal dynamics using a {\em chaotic
spin-glass shard} Hamiltonian \cite{gs,jose}.  As a main feature, the
chaotic bath model assumes that {\em no} integrals of motion exist for
the bath other than the energy.  This differs from the
dipolarly-coupled environment model (and the real FAp sample), where
the environment chains are similar to the central chain and, in the
absence of pulses, the total magnetization of the central chain and
the bath is conserved.

\begin{figure}[t]
\vspace*{5mm}
\includegraphics[scale=0.2,angle=-0]{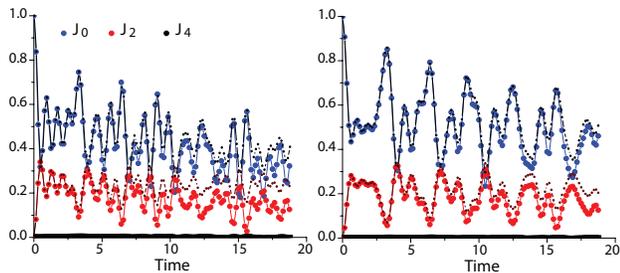}
\caption{(Color online) $J_0(t)$, $J_2(t)$, and $J_4(t)$ for a chain
with thermal initial state (left), and for the chain with
end-polarized initial state (right). The chain is coupled to the bath
of nine spins with chaotic glass-shard internal dynamics described by
Eq. (\ref{glass}). Dash-dotted lines: Simulations with the
dipolarly-coupled spin environment described in
Sec.~\ref{sec:ChainPlusBath}.}
\label{figA2:chaoticbath}
\end{figure}

Specifically, in our case we choose the chain-bath coupling to mimic
the arrangement of FAp samples: each chain spin is coupled to six bath
spins, the coupling has a homonuclear secular dipolar form, similar to
Eq.~(\ref{eq:dip}), and the coupling constants $b_{j\ell}$ for each
pair of a chain and a bath spin are drawn uniformly from the interval
$[-\sqrt{3}\cdot 0.025, \sqrt{3}\cdot 0.025]$.  This ensures that the
rms coupling between one bath spin and one chain spin is equal to the
experimental value $b_\times/b\approx 0.025$, see
Eq.~(\ref{eq:bcross}).  Nine bath spins are located on a $3\times 3$
square lattice, with a Hamiltonian
\begin{equation}
H_B = \sum_{\langle k,l\rangle} \Gamma_{kl} S^x_k S^x_l + \sum_k h^z_k
  S^z_k + \sum_k h^x_k S^x_k, 
\label{glass}
\end{equation}
\noindent
where the summation in the first term is over NN pairs.  The random
couplings $\Gamma_{kl}$ and the local magnetic fields $h^{x,z}_k$ are
drawn uniformly from the intervals $[-\Gamma_0,\Gamma_0]$ and
$[-h_0,h_0]$, respectively, with the values of $\Gamma_0$ and $h_0$
adjusted to ensure: (i) chaotic regime, and (ii) correct
characteristic energies for the spin dynamics inside the bath
To achieve the latter, note that for a FAp chain with NN couplings
only, and $N\gg 1$ spins, $\tr{H^2} = (6/16)\ N\ \tr{\openone}$, so
that the rms energy per spin is 6/16. Correspondingly, the values of
$\Gamma_0$ and $h_0$ were adjusted to give approximately the same rms
energy per spin.

The results of the simulations for the thermal initial state and for
the end-polarized initial state are given in
Fig.~\ref{figA2:chaoticbath}.  It is clearly seen that the interaction
with the bath leads to significant damping of the oscillations of
$J_0(t)$ and $J_2(t)$, and to an overall decay of these coherences,
although the mirror time remains clearly visible.  Interestingly, as
also noted in the text, the decay of the zero- and second-order
coherences $J_0(t)$ and $J_2(t)$ for the end-polarized state is slower
than for thermal state. To further appreciate this, we compare the
dynamics of $J_0(t)$ and $J_2(t)$ for the two bath models we examined
in Fig.~\ref{figA2:chaoticbath}.  The $J_0(t)$ signals for both bath
models stay close to each other, while exhibiting significant damping
of oscillations and overall decay in comparison with the analytical
results for the isolated chain. In contrast, $J_2(t)$ for the random
dipolarly-coupled environment stays rather close to the analytical
prediction for the isolated chain, whereas $J_2(t)$ for the chaotic
bath decays in the same way as $J_0(t)$ does.  This suggests that {the
presence of extra integrals of motion does not significantly affect
the dynamics of $J_0(t)$, whereas higher-order MQCs might more
sensitively depend upon details of the open-system dynamics.



\begin{thebibliography}{99}
\bibitem{AHTnote} The relevant $16$-pulse sequence {\tt S} may be
understood starting from a simpler $2$-pulse cycle {\tt C} which also
simulates the DQ Hamiltonian, along with its time-reversed version
$\bar{\tt C}$.  The primitive pulse cycle ${\tt C}=[\frac{\Delta}{2} X
\Delta' \frac{\Delta}{2}]$, where $\Delta'= 2\Delta + w$, $\Delta, w$
being the pulse delay and the pulse width, respectively.  Then ${\tt
S}={\tt C}\cdot \bar{\tt C}\cdot \bar{\tt C}\cdot {\tt C}\cdot
\bar{\tt C} \cdot {\tt C} \cdot {\tt C} \cdot \bar{\tt C}$
\cite{Ramanathan03}.
\bibitem{comment} As it turns out, the peak at the mirror time is no
longer the second largest for the thermal state in the presence of
cross-chain coupling, whereas it is still such for the end-polarized
state.
\end{thebibliography}
\end{document}